\documentclass[preprint]{aastex}

\newcommand{\Spitzer}{{\sl Spitzer}}

\newcommand{\Msun}{\mbox{$M_{\sun}$}}

\newcommand{\Lsun}{\mbox{$L_{\sun}$}}

\newcommand{\Mjup}{\mbox{$M_{\rm Jup}$}}

\newcommand{\etal}{et al.}
\newcommand{\eg}{e.g.}
\newcommand{\ie}{i.e.}
\newcommand{\cf}{cf.}

\newcommand{\Zsun}{\mbox{$Z_{\odot}$}}
\newcommand{\htwoo}{{\hbox{H$_2$O}}}   
\newcommand{\htwo}{{\hbox{H$_2$}}}     
\newcommand{\meth}{{\hbox{CH$_4$}}}   

\newcommand{\Lbol}{\mbox{$L_{\rm bol}$}}
\newcommand{\Teff}{\mbox{$T_{\rm eff}$}}
\newcommand{\logg}{\mbox{$\log(g)$}}

\newcommand{\hdobject}{\hbox{HD~3651B}}
\newcommand{\globject}{\hbox{Gl~570D}}

\slugcomment{submitted Sept 12, 2006 to ApJ; accepted December 20, 2006}

\shorttitle{HD 3651B: A Benchmark Late-T Dwarf}
\shortauthors{Liu \etal}

\begin{document}

\title{The Late-T Dwarf Companion to the Exoplanet Host Star HD~3651:\\A
  New Benchmark for Gravity and Metallicity Effects in Ultracool Spectra}


\author{Michael C. Liu,\altaffilmark{1,2,3}
        S. K. Leggett,\altaffilmark{4}
        Kuenley Chiu\altaffilmark{5}}

\altaffiltext{1}{Institute for Astronomy, University of Hawai`i, 2680
  Woodlawn Drive, Honolulu, HI 96822; mliu@ifa.hawaii.edu}
\altaffiltext{2}{Alfred P. Sloan Research Fellow}
\altaffiltext{3}{Visiting Astronomer at the Infrared Telescope Facility,
  which is operated by the University of Hawaii under Cooperative
  Agreement no. NCC 5-538 with the National Aeronautics and Space
  Administration, Science Mission Directorate, Planetary Astronomy
  Program.}
\altaffiltext{4}{Gemini Observatory Northern Operations Center, 670 North 
A'ohoku Place, Hilo, HI 96720}
\altaffiltext{5}{Astrophysics Group, School of Physics, University of 
Exeter, Stocker Road, Exeter, EX4 4QL, UK}

\begin{abstract}
We present near-IR (1.0--2.5~\micron) photometry and spectroscopy
of \hdobject, the low-luminosity, wide-separation (480~AU) companion to
the K0V exoplanet host star HD~3651A.
We measure a spectral type of T7.5$\pm$0.5 for \hdobject, confirming
both its substellar nature and the fact that wide-separation brown
dwarfs and giant planets can co-exist around the same star.
We estimate an age of 3--12~Gyr for the primary star HD~3651A and find
that it is $\approx$3$\times$ older than the K4V star Gl~570A
($\approx$1--5~Gyr), the host star of the T7.5 dwarf Gl~570D.
We derive a bolometric luminosity of $\log(\Lbol/\Lsun)=-5.58\pm0.05$
for \hdobject\ and infer an effective temperature of 780--840~K and a
mass of 40--72~\Mjup; the luminosity and temperature are among the
lowest measured for any brown dwarf.
Furthermore, \hdobject\ belongs to the rare class of substellar objects
that are companions to main-sequence stars and thus provides a new
benchmark for studying very low-temperature objects.
Given their similar temperatures ($\Delta\Teff\approx30$~K) and
metallicities ($\Delta$[Fe/H] $\approx$ 0.1~dex) but different ages, a
comparison of \hdobject\ and \globject\ allows us to examine
gravity-sensitive diagnostics in ultracool spectra.
We find that the expected signature of HD~3651B's higher surface gravity
due to its older age, namely a suppressed $K$-band flux relative to
\globject, is not seen.  Instead, the $K$-band flux of \hdobject\ is
enhanced compared to \globject, indicative of a younger age.
Thus, the relative ages derived from interpretation of the T~dwarf
spectra and from stellar activity indicators appear to be in discord.
One likely explanation is that the $K$-band fluxes are also very
sensitive to metallicity differences.  Metallicity variations may be as
important as surface gravity variations in causing spectral differences
among field late-T dwarfs.
\end{abstract}



\section{Introduction}

Discovery and scrutiny of brown dwarfs have been fertile avenues for
understanding self-luminous objects, extending traditional stellar
astrophysics into new domains of mass and effective temperature.  The
coolest brown dwarfs, the T~dwarfs, are characterized by very red
optical colors from pressure-broadened alkali resonance lines and very
blue near-IR colors from strong methane absorption and collision-induced
\htwo\ absorption \citep[e.g.][]{1995Sci...270.1478O, geb01, kirk05}.
T~dwarfs are the lowest luminosity and coolest objects directly detected
outside of our solar system, with bolometric luminosities (\Lbol) of
$\lesssim10^{-4.5}\Lsun$ and effective temperatures (\Teff) of
$\approx700-1300$~K \citep[e.g.][]{2004AJ....127.2948V, gol04,
  2006ApJ...639.1095B}.  As such, analyzing their spectra to infer
temperature, gravity, metallicity, and mass is a key pathway to
understanding the properties of gas-giant extrasolar planets.

The first T~dwarf, Gl~229B, was discovered as a 45-AU companion around a
nearby M~dwarf \citep{1995Natur.378..463N}.  However, since then the
vast majority have been found as free-floating field objects by the
2MASS and SDSS wide-field surveys \citep[e.g.][]{1999ApJ...522L..65B,
  2000ApJ...536L..35L, chiu05}, with about 100 T~dwarfs identified to
date.  Such a populous sample has been a boon for probing substellar
astrophysics.  But this sample is also inevitably hindered by the
unknown ages and metallicities of the field population.
Binary brown dwarfs provide a partial solution to this challenge, as
they constitute systems of common age and metallicity; however, the
absolute values of these quantities are unknown for binaries.  
In this regard, brown dwarfs that are resolved companions to stars are
highly prized, as their distances, ages and metallicities can be
established from their primary stars, given the conservative assumption
that the systems formed coevally and from material of the same
composition.
These brown dwarf companions serve as ``benchmarks'' for studying
substellar atmospheres and evolution.
Such objects are rare, as numerous published (and unpublished) imaging
surveys have searched for brown dwarf companions to nearby main-sequence
stars with very limited success \citep[e.g.][]{lloyd00,
  2001AJ....121.2189O, 2004AJ....127.2871M, 2006AJ....132.1146C}.

\citet{2006astro.ph..8484M} have recently identified a low-luminosity
companion to the nearby K0V star HD~3651 (GJ~27, 54~Psc, HR~166,
HIP~3093, SAO~74175).  The primary star is also notable as it possesses
a sub-Saturn mass planet ($M \sin i = 0.20$~\Mjup) with an orbital
semi-major axis of 0.3~AU \citep{2003ApJ...590.1081F}.  \hdobject\ has a
projected separation of 43\arcsec\ (480~AU) and is confirmed to be
physically associated by its common proper motion with HD~3651A.  Given
the distance of 11.11$\pm$0.09~pc to the primary
\citep{1997A&A...323L..49P}, the very faint absolute $H$-band magnitude
and the absence of an optical counterpart in photographic plates suggest
that HD~3651B is a very cool brown dwarf.

In this paper, we present near-IR photometry and spectroscopy
to characterize \hdobject\ and compare it to Gl~570D, the coolest known
companion around a main-sequence star.  After our paper was submitted,
\citet{2006astro.ph..9464L} reported identification and characterization
of \hdobject\ based on \Spitzer/IRAC and ground-based near-IR
measurements, and \citet{2006astro.ph..9556B} reported near-IR
spectroscopy and analysis.


\section{Observations}

\subsection{Photometry}

We obtained near-infrared (IR) imaging of \hdobject\ on
2006~September~3~UT from the United Kingdom Infrared Telescope (UKIRT)
located on Mauna Kea, Hawaii.  We used the facility IR camera UFTI
\citep{2003SPIE.4841..901R} and the $J$ (1.25~\micron), $H$
(1.64~\micron), and $K$ (2.20~\micron) filters from the Mauna Kea
Observatories (MKO) filter consortium \citep{mkofilters1, mkofilters2}.
We read out a 512 $\times$ 512 pixel region of UFTI's detector centered
on \hdobject, leading to a 47\arcsec\ field of view.  The primary star
HD~3651A was placed off the array.
UKIRT has a fast-steering secondary mirror that provides tip-tilt
correction, producing an image quality of 0.55\arcsec\ FWHM during our
observations.  \hdobject\ appeared as a single object in the images.
Sky conditions were photometric.
We obtained a series of 9 dithered images in each filter, for a total
on-source integration time of 9~minutes in $J$, $H$, and $K$ each.  The
images were reduced in a standard fashion using the facility data
pipeline.
The flux calibration was established from observations of the UKIRT
standard stars FS~154 and FS~102 \citep{2006MNRAS.tmp.1209L}, observed
immediately before and after \hdobject\ and at similar airmass.  Table~1
presents our final UKIRT/UFTI photometry.

\subsection{Spectroscopy}

We obtained spectroscopy of \hdobject\ on 2006~September~4 and~5~UT
using the UKIRT facility spectrograph CGS4 \citep{1993SPIE.1946..547W}.
CGS4 has a Santa Barbara Research Center (SBRC) 256$\times$256 pixel
InSb detector.  The two-pixel (1.2\arcsec) slit was used for all
observations.  \hdobject\ was observed at three grating settings to span
the $J$-, $H$- and $K$-bands, covering 1.03--1.35, 1.38--2.02 and
1.88--2.52~\micron\ with spectral resolutions of 21~\AA\ at $J$
($R=\lambda/\Delta\lambda\approx600$) and 50~\AA\ at $H$ and $K$
($R\approx330$ and 440).  Individual exposure times were 120~seconds for
the $J$ setting, 60~seconds for $H$ and 40~seconds for $K$, with total
on-source exposure times of 48, 32, and 59~minutes at $J$, $H$, and $K$,
respectively.  The target was nodded along the slit by 60\arcsec\ in
order to avoid scattered light from the very bright primary star.
The spectrum of the target at $\lambda < 1.2$~\micron\ was still
difficult to extract due to the light of the primary, and thus we do not
use it.
CGS4 has a calibration unit with lamps that provide accurate
flat-fielding and wavelength calibration.
The F6V star HD~615 was used as a calibrator to remove the effects of
the terrestrial atmosphere, with H~I recombination lines in its spectrum
removed artificially prior to ratioing.
The three separate spectra obtained for \hdobject\ were merged together
based on our UKIRT/UFTI photometry from the night before.  After
scaling, the 1.94--2.01~\micron\ regions that overlapped in the $H$ and
$K$-band spectra agreed to about $5\times10^{-18}$
W/m$^2$/\micron\ (\cf, the $K$ and $H$-band emission peaks of
$2.1\times10^{-16}$ and $8.2\times10^{-16}$ W/m$^2$/\micron,
respectively), thereby indicating the robustness of the merging process.

We also obtained low-resolution ($R\approx$150) spectra of \hdobject\ on
2006~September~13~UT from NASA's Infrared Telescope Facility (IRTF)
located on Mauna Kea, Hawaii.  Conditions were photometric with
excellent seeing conditions, around 0.5\arcsec\ FWHM at $K$-band near
zenith.  We used the facility near-IR spectrograph Spex
\citep{1998SPIE.3354..468R} in prism mode, obtaining
0.8--2.5~\micron\ spectra in a single order.  We used the
0.5\arcsec\ wide slit, oriented at the parallactic angle to minimize the
effect of atmospheric dispersion.  \hdobject\ was nodded along the slit
in an ABBA pattern, with individual exposure times of 200~sec, and
observed over an airmass range of 1.00--1.08.  The telescope was guided
using images of the nearby star 2MASS~J00391738+2115104 obtained with
the near-IR slit-viewing camera.  The total on-source exposure time was
73~min.  We observed the A0~V star HD~7215 contemporaneously with
\hdobject\ for flux and telluric calibration.  All spectra were reduced
using version 3.3 of the SpeXtool software package
\citep{2003PASP..115..389V,2004PASP..116..362C}.  The reduced IRTF/Spex
spectrum is plotted in Figure~\ref{fig:plot-spectra}, in good agreement
with the UKIRT/CGS4 spectrum.  Also, the IRTF/Spex data provide the
$\lambda\lesssim1.2~\micron$ region that was unobtainable with the UKIRT
data.


\section{Results}

\subsection{Near-IR Spectral Types and Photometry}

\begin{figure}
\centerline{\includegraphics[height=7.5in,angle=0]{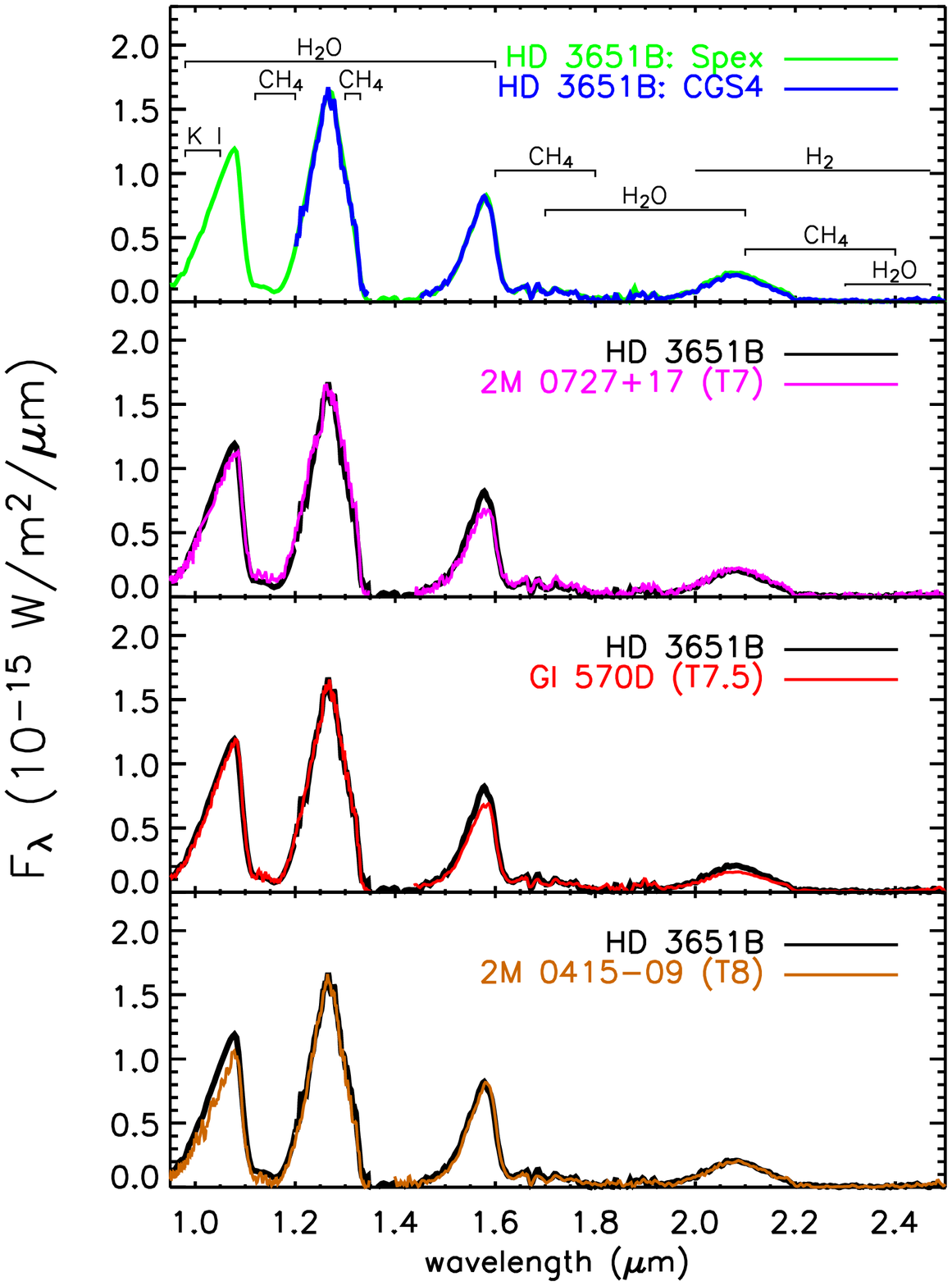}}
\caption{\normalsize {\bf Top panel:} Our near-IR spectra of
  \hdobject\ obtained with UKIRT/CGS4 and IRTF/Spex.  {\bf Other
    panels:} Same spectra of \hdobject\ plotted as a thick black line,
  with mostly the UKIRT/CGS4 data plotted and the IRTF/Spex data filling
  in the $<$1.2~\micron\ and 1.35--1.45~\micron\ regions.  Spectra of
  other late-T dwarfs are plotted as colored lines.  Their near-IR spectra
  ($>$1~\micron) were also obtained with CGS4, with the same intrumental
  setup and spectral resolution \citep{2001ApJ...556..373G, geb01,
    2004AJ....127.3553K}, and the $<$1~\micron\ spectra come from
  \citet{2003ApJ...594..510B}.  The spectra have been normalized by
  their peak flux.\label{fig:plot-spectra}}
\end{figure}

Figure~\ref{fig:plot-spectra} presents our near-IR spectra of \hdobject,
showing the strong water and methane absorption bands that are the
hallmarks of the T~spectral class.  Spectra of other late-T~dwarfs are
shown for comparison.
We classified \hdobject\ from the system of five spectral indices
established by \citet{2005astro.ph.10090B}, as measured independently
from our UKIRT/CGS4 and IRTF/Spex data (Table~2).  The UKIRT/CGS4 data
indicate a spectral type of T7.5, while the IRTF/Spex data indicate T8.
(The difference does not arise from the differing spectral resolution of
the two datasets, as we verified by smoothing the CGS4 data.)
Figure~\ref{fig:plot-spectra} shows that the IRTF/Spex data slightly
suggest a later type, based on the very slightly deeper \htwoo\ and
\meth\ absorption.  However, in both datasets, \hdobject\ does not
appear as late as the T8~dwarf 2MASS~0415$-$0935 based on the depth of
the 1.15~\micron\ \htwoo\ absorption band.
We also visually classified \hdobject\ by comparing with UKIRT/CGS4 and
IRTF/Spex spectra of late-T dwarfs classified by Burgasser \etal, which
have the same resolutions and instrumental setups as our data.  For data
from each instrument, the depth of the \htwoo\ and \meth\ absorption
bands were examined, normalizing the spectra of \hdobject\ and the
comparison objects to their peak fluxes in the $J$, $H$, and $K$-bands.
This process confirmed that \hdobject\ is later than T7 but earlier than
T8.  In fact, the depth of the absorption bands for \hdobject\ as judged
by the indices and by eye are nearly identical to the T7.5~dwarf
\globject.  

Therefore, we assign a spectral type of T7.5 for \hdobject, with the
nominal error of $\pm$0.5 subclasses from the Burgasser \etal\ system.
Our typing is in agreement with other results.
\citet{2006astro.ph..8484M} estimate T7--T8 based solely on the $H$-band
absolute magnitude.  Using independent sets of IRTF/Spex prism spectra,
\citet{2006astro.ph..9464L} determine T7.5$\pm$0.5 based on visual
examination, and \citet{2006astro.ph..9556B} find T8$\pm$0.5 based on
spectral index measurements.  With such a late-type spectrum,
\hdobject\ is unambiguously a substellar object.

\begin{figure}
\includegraphics[width=5in,angle=90]{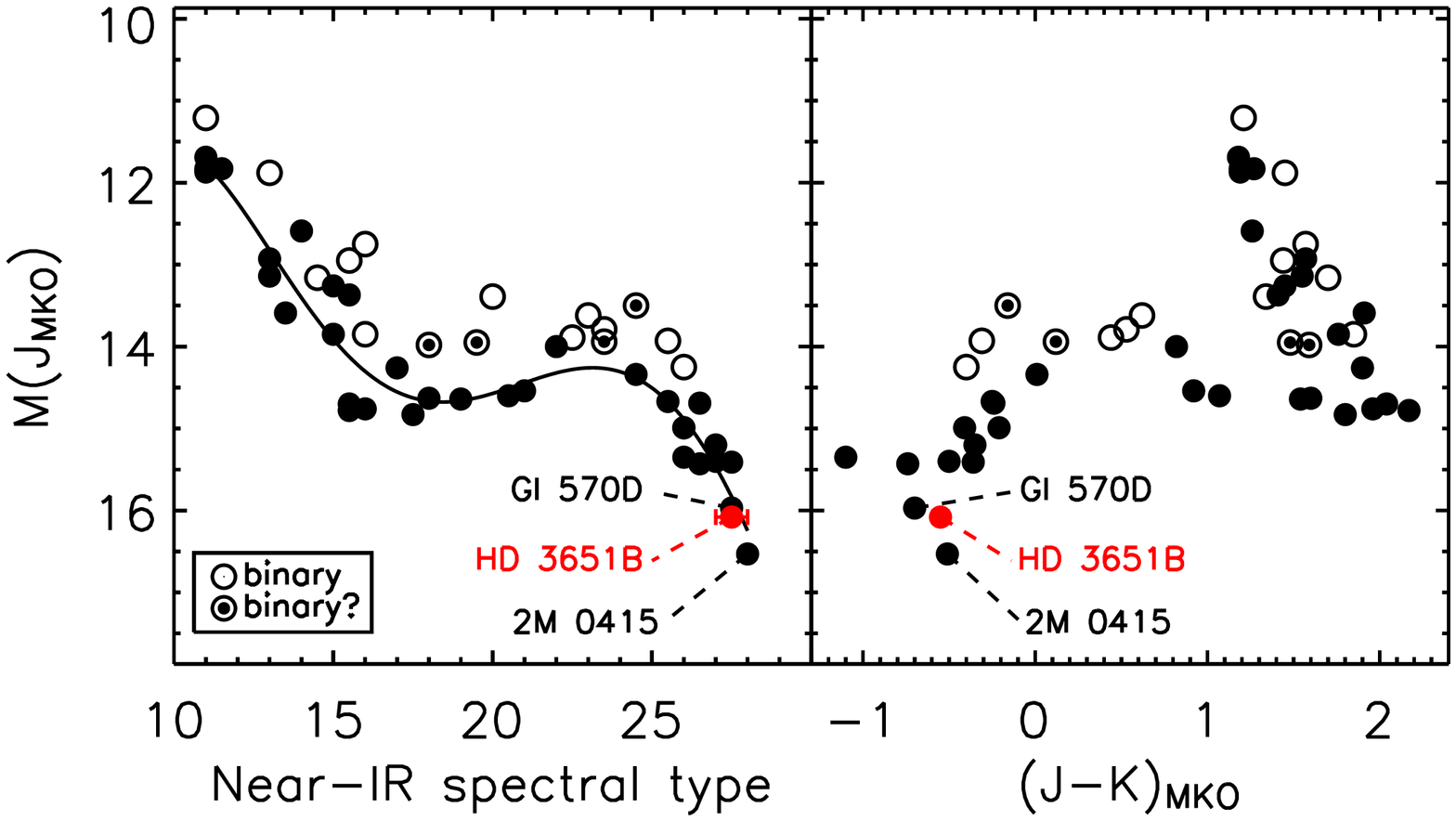}
\vskip -6ex
\caption{\normalsize Near-IR photometry of \hdobject\ compared to other
  L~and T~dwarfs with known distances, largely from
  \citet{2002AJ....124.1170D} and \citet{2004AJ....127.2948V} as
  tabulated by \citet{2004AJ....127.3553K}.  Known binaries are plotted
  as open circles, and possible binaries as encircled dots.  See
  \citet{2006astro.ph..5037L} for additional references and discussion
  of possible binaries.  The photometric measurements use the MKO
  system, and their errors are comparable to or smaller than the symbol
  size.  {\bf Left:} $J$-band absolute magnitude as a function of
  near-IR spectral type and $J-K$ color.  Spectral types are based on
  the \citet{geb01} scheme for the L~dwarfs and the
  \citet{2005astro.ph.10090B} scheme for the T~dwarfs, with L0 being 10,
  L1 begin 11, T0 being 20, etc. on the x-axis.  The solid line shows a
  5th-order polynomial fit, excluding known and possible binaries and
  \hdobject.  The T7.5~dwarf Gl~570D and the T8~dwarf 2MASS~0415$-$0935
  are labeled. {\bf Right:} $J$-band absolute magnitude versus $J-K$
  color. \label{fig:plot-sptype}}
\end{figure}

Figure~\ref{fig:plot-sptype} shows our near-IR photometry of
\hdobject\ compared to other ultracool dwarfs; again, \hdobject\ appears
to be very similar to \globject.  Our UKIRT/UFTI $H$ and $K$-band
photometry for \hdobject\ agrees within the errors with photometry
obtained by \citet{2006astro.ph..9464L} using the IRTF/Spex slit-viewing
camera.  However, the $J$-band results differ by 0.15~mag, in that our
UKIRT/UFTI measurement two months later is fainter.
To explore this discrepancy, we used our single-order IRTF/Spex spectrum
to synthesize the near-IR colors and found conflicting results: our
Spex-synthesized $J-H$ agrees with our UKIRT/UFTI photometry but the
synthesized $H-K$ is redder by 0.08~mag.
It is likely that the discrepancy with the Luhman \etal\ $J$-band
photometry is not significant at these levels.  
The IRTF/Spex slit-viewing camera is used primarily for acquisition and
guiding, and it has not been rigorously tested for precision photometry,
e.g. the linearity of the detector response has not been
well-characterized (J. Rayner, priv. comm.).
However, it is also
possible that \hdobject\ is variable at the $\approx$10\% level ---
further monitoring could be valuable.

Similarly, the $H-K$ discrepancy between our IRTF/Spex and UKIRT
datasets is within the plausible errors of the overall flux calibration
for the Spex data.  \citet{2006ApJ...639.1095B} examined the consistency
of broad-band photometric colors compared to colors synthesized from
low-resolution IRTF/Spex spectra of 16~late-T dwarfs.  They found
typical deviations between the observed and synthesized colors of 5\% or
less, with a few sources having differences as large as 15\%.

\subsection{Age of the HD~3651AB System}

We consider several methods for establishing the age of the primary star
HD~3651A.  Age determination for main-sequence field stars is
challenging and imperfect.
For solar-type stars, methods for estimating ages largely rely on the
increase in stellar rotation period as stars grow older.  Stars spin
down as they age because stellar winds carry away angular momentum; the
increased rotation periods then lead to a decline in stellar activity
due to the underlying stellar dynamo.

For solar-type stars, chromospheric activity as traced by CaII HK
emission provides an age estimate.  \citet{don93, 1998csss...10.1235D}
provide an age calibration for this index:
\begin{equation}
\log(t) = 10.725 - 1.334 R_5 + 0.4085 R_5^2  - 0.0522 R_5^3
\end{equation}
where $R_5 = 10^5 R\arcmin_{HK}$, valid for $\log(R\arcmin_{HK}) $ = 
$-$4.25 to $-$5.2.
\citet{2003AJ....126.2048G} measure $\log(R\arcmin_{HK}) = -5.09$ for
HD~3651A with an uncertainty of about 0.05 (R. Gray, priv. comm.) and
describe it as an inactive star.  \citet{2004ApJS..152..261W} find
$\log(R\arcmin_{HK}) = -5.02$, with a full range of about 10\% over
7~years of measurements. (\citealp{1991ApJS...76..383D} report a 40\%
change in CaII~HK emission over 16 years of monitoring.)
These measurements lead to an age estimate of about 5--9~Gyr.  An error
estimate is not available for this technique.  However, out of a sample
of 21 binaries studied by \citet{1998csss...10.1235D}, most had age
estimates differing by $\lesssim$2~Gyr.

X-ray emission of solar-type stars also declines with age.
\citet{1995A&A...294..515H} measure $L_X/L_{bol} = -5.69$ for HD~3651A
with a 30\% uncertainty.  \citet{1998PASP..110.1259G} provides an age
calibration based on scaling relations for stellar activity:
\begin{equation}
\log(L_X/L_{bol}) = -6.38 - 2.6\alpha\log(t_9/4.6) + \log[1 + 0.4(1-t_9/4.6)]
\end{equation}
where $t_9$ is the age in Gyr and $\alpha$ is the coefficient that
relates rotation period to stellar age, either $\alpha=0.5$
\citep{1972ApJ...171..565S} or $\alpha = 1/e$ \citep{wal91}.  Following
\citet{wils01}, we adopt the zeropoint of $-$6.38 based on the X-ray
luminosity of the Sun from \citet{1987ApJ...315..687M}.  Including the
uncertainties, we estimate an age of 0.9--2.2~Gyr for HD~3651A.
As a point of reference, the X-ray luminosity of HD~3651A is
$\log(L_X)=27.6$ \citep{1995A&A...294..515H}.  This is about 6~times
fainter than Hyades stars of similar spectral type
\citep{1995ApJ...448..683S}.  
\citet{2005ApJS..160..390P} estimate $L_X \propto t^{-0.77}$ for
solar-type stars, implying that HD~3651A is about 4$\times$ older than
the Hyades (650~Myr) or around 3~Gyr.

Due to angular momentum carried away by stellar winds, the rotation
periods of solar-type stars increase as they age, believed to follow a
power-law relation of $P_{rot} \sim t^{\alpha}$
\citep[e.g.][]{1972ApJ...171..565S}, where $\alpha$ is the same as in
Equation~2.  \citet{1983ApJ...275..752B} report a period of 48~d for
HD~3651A's chromospheric activity, which we adopt as the stellar
rotation period.  Using the Sun as a reference point ($P_{rot} = 26$~d
and $t=4.6$~Gyr), the scaling relation gives an implausibly large age of
16--24~Gyr.  \citet{1999A&A...348..897L} have provided an age
calibration for main-sequence stars based on a sample from the {\sl
  Hipparcos} catalog:
\begin{equation}
\log(t_9) = 2.667 \log(P) - 0.944 (B-V) - 0.309 [{\rm Fe/H}] + 6.530 .
\end{equation}
where $t_9$ is the age in Gyr and $P$ is the period in days.  With
$B-V=0.85$~mag \citep{1997A&A...323L..49P} and [Fe/H]=0.09--0.16
\citep{2003AJ....126.2048G, 2004A&A...415.1153S,2005ApJS..159..141V},
this gives an age of 15~Gyr.  \citet{1999A&A...348..897L} caution that
their relation is less accurate for stars older than 3~Gyr.

Finally, from high resolution spectroscopic analysis combined with
bolometric magnitudes and theoretical stellar evolutionary isochrones,
\citet{2005ApJS..159..141V} derive an age estimate of 8.2~Gyr with a
possible range of 3--12~Gyr, and \citet{2006astro.ph..7235T} infer a
minimum age of 11.8~Gyr.  The quoted age range from Valenti \& Fischer
spans the aforementioned activity-based estimates from the X-ray data
($\approx$1--3~Gyr) and CaII HK data ($\approx$5--9~Gyr), and thus we
adopt an age of 3--12 for HD~3651A, with a geometric mean of 6~Gyr.
This old age is supported by the star's slow rotation and the Takeda
\etal\ estimate.

\subsection{Luminosity, Mass, and Effective Temperature}

To detemine the bolometric luminosity of \hdobject, we use a $K$-band
bolometric correction ($BC_K$) of 2.07$\pm$0.13~mag, based on the
\citet{gol04} polynomial relation of $BC_K$ versus near-IR spectral
type.  This gives $\log(\Lbol/\Lsun)=-5.58 \pm 0.05$, with the
uncertainty coming from the quadrature sum of the uncertainties in the
$K$-band absolute magnitude, the distance to the system, the $BC_K$ due
to the spectral typing uncertainty, and the scatter about the fitted
polynomial relation.  \hdobject\ has the second smallest \Lbol\ measured
among all brown dwarfs with trigonometric distances, comparable to that
of the T7.5~dwarf Gl~570D and exceeded only by the T8~dwarf
2MASS~0415$-$0935, which have $\log(\Lbol/\Lsun)=-5.53 \pm 0.05$ and
$-5.73\pm0.07$, respectively \citep{gol04}.\footnote{Analysis of the
  near-IR spectra of the T8~dwarf 2MASS~0939$-$2448 and the T7.5~dwarf
  2MASS~1114$-$2618 by \citet{2006ApJ...639.1095B} suggest that these
  two objects may be even cooler and lower luminosity than
  2MASS~0415$-$0935, though they do not yet have parallaxes
  measurements.}

Instead of the $BC_K$ from the polynomial relation as a function of
spectral type, we could also have used the individual $BC_K$ values
determined for the four late-T dwarfs in the Golimowski \etal\ (2004)
sample: 2MASS~0727+1710 (T7; $BC_K=2.24\pm0.13$~mag), Gl~570D (T7.5;
1.90$\pm$0.13~mag); 2MASS~1217$-$0311 (T7.5; 2.28$\pm$0.13~mag); and
2MASS~0415$-$0935 (T8; 2.03$\pm$0.13~mag), where the spectral types are
those assigned by \citet{2005astro.ph.10090B}.  The average of the two
T7.5 dwarfs is 2.09~mag, consistent with our adopted value.  If we adopt
the Gl~570D value based on its close spectral resemblance to HD~3651B,
we would obtain an \Lbol\ value 0.08~dex brighter, which would not have
a significant impact on our conclusions.  Finally,
\citet{2006astro.ph..9464L} used near-IR spectra, \Spitzer/IRAC thermal
IR (3.2--9.2~\micron) photometry, and an assumed long wavelength
Rayleigh-Jeans distribution to derive a bolometric luminosity of
$\log(\Lbol/\Lsun)=-5.60 \pm 0.05$, in excellent agreement with our
assessment.

To estimate the \Teff\ and mass of \hdobject, we use the observational
constraints of the derived \Lbol\ and estimated age of 6$_{-3}^{+6}$~Gyr
combined with the solar-metallicity evolutionary models of
\citet{1997ApJ...491..856B}.
For a nominal age of 6~Gyr, we find an effective temperature of 810~K
and a mass of 56~\Mjup.  The ranges of these values are 780--840~K and
masses of 40--72~\Mjup, with younger assumed ages leading to cooler
temperatures and lower masses.  The resulting derived properties are
given in Table~1.\footnote{In principle, one should use evolutionary
  models computed for the same metallicity as the parent stars.
  However, the slightly super-solar metallicity of HD~3631A (discussed
  in \S~4) should not lead to a significant difference in the derived
  properties.  Tables 2 and~3 of \citet{2006ApJ...647..552S} show that
  for late-T dwarfs, changing the metallicity from 0.0~dex to 0.3~dex
  changes the model-derived mass by $\lesssim$5\%, \Teff\ by
  $\lesssim$1\%, and \logg\ by $\lesssim$0.05~dex for a given age.}

\section{Discussion}

\subsection{Relative Ages, Masses and Temperatures of HD~3651B and Gl~570D}

HD~3651B is remarkably similar to the T7.5 dwarf \globject, a
wide-separation companion to the triple system Gl~570ABC
\citep{2000ApJ...531L..57B, 2001ApJ...556..373G}, separated by 1525~AU
from the K4V primary star.  These two brown dwarfs have very comparable
properties, namely their near-IR spectra (both spectral type T7.5),
$JHK$ absolute magnitudes (HD~3651B is $\lesssim$0.1~mag fainter), and
$JHK$ colors (HD~3651B is redder by $\lesssim$0.15~mag).  Their primary
stars both have about solar metallicity, [Fe/H] = 0.01--0.10 for Gl~570A
\citep{1997A&AS..124..299C, 1998A&AS..129..237F, 2005A&A...437.1127S,
  2005ApJS..159..141V} and [Fe/H] = 0.09--0.16 for HD~3651A
\citep{2003AJ....126.2048G, 2004A&A...415.1153S,2005ApJS..159..141V}.
Hence, we would expect that differences between these two brown dwarfs
arise primarily from their differing masses and ages, whose
observational manifestation is surface gravity.
\citet{2001ApJ...556..373G} estimated an age of 2--5~Gyr for Gl~570A,
younger than our estimate of 3--12~Gyr for HD~3651A, but the two
estimates derive from different methods.  Here we re-examine the
relative ages of these two primary stars and in the next section
consider the implications in interpreting the spectra of their late-T
dwarf companions.

Gl~570A is more chromospherically active as judged by its
$\log(R\arcmin_{HK})$ values $-$4.49 and $-$4.75
\citep{1996AJ....111..439H, 1991ApJ...375..722S}, compared to
$\log(R\arcmin_{HK}) \approx -5.05$ for HD~3651A (\S~3.2).
Equation~1 gives corresponding ages of 0.8 and 2.2~Gyr for Gl~570A,
compared to $\approx$7~Gyr for
HD~3651A.\footnote{\citet{2000A&AS..142..275S} report
  $\log(R\arcmin_{HK}) = -4.21$ for Gl~570A, which is higher level of
  activity than covered by the \citet{1998csss...10.1235D} calibration,
  suggesting an age of $\lesssim$10~Myr.  However, the absence of Li~I
  6708~\AA\ absorption indicates that Gl~570A is at least a zero-age
  main sequence star \citep{2002A&A...384..912R}.}
\citet{2001ApJ...556..373G} suggested that the young age inferred from
the CaII HK data may be due to the fact that Gl~570A was observed during
a period of enhanced activity.
As a point of reference, the Sun varies from $\log(R\arcmin_{HK}) =
-5.10$ to $-$4.75, which would lead to chromospherically inferred ages
of 3--8~Gyr \citep[e.g.][]{1998ASPC..154..153B}.  Assuming the CaII HK
variability of Gl~570A and HD~3651A is comparable to the solar cycle,
the two stars would have to had been observed at nearly the opposite
extremes of their activity cycles to still have the same age.  We
discount this possibility and conclude that the chromospheric data
indicate that Gl~570A is younger than HD~3651A.

Similarly, the other age indicators we have considered support a younger
age for Gl~570A compared to HD~3651A.  Based on $\log(L_X)=27.7$ from
\citet{2004A&A...417..651S} and bolometric corrections from
\citet{1995ApJS..101..117K}, we find Gl~570A has
$\log(L_X/\Lbol)=-5.27$.  Using Equation~2, this gives an age of
0.4--0.8~Gyr, compared to 1--3~Gyr inferred for HD~3651A from the same
approach.  As pointed out by \citet{2001ApJ...556..373G}, Gl~570A has a
lower X-ray luminosity compared Hyades stars of similar spectral type,
setting a lower age limit of 650~Myr.
Also, Gl~570A has a somewhat shorter rotation period of 40~days,
compared to 48~days for HD~3651A, implying a $\approx$50--60\% younger
age for Gl~570A based on Equation~3 (\S3.2).  Finally, isochrone
analyses for Gl~570A by \citet{2005ApJS..159..141V} and
\citet{2006astro.ph..7235T} give ages of 3.3$_{-3.1}^{+8.3}$ Gyr and
$<$0.6~Gyr, respectively.

Table~3 summarizes the age estimates for HD~3651A and Gl~570A.  While
there is considerable scatter in the absolute ages, the data all suggest
that Gl~570A is $\approx$3$\times$ younger than HD~3651A.  In the
analysis that follows, we adopt a conservative age range of 1--5~Gyr for
Gl~570D, with a geometric mean of 2~Gyr.  The upper limit of 5~Gyr is
the same as that of \citet{2001ApJ...556..373G} and derives from the
scatter of the \citet{1998csss...10.1235D} calibration sample.  Our
adopted lower age limit of 1~Gyr is younger than the 2~Gyr used by
\citet{2000ApJ...531L..57B} and \citet{2001ApJ...556..373G}.  Their
value is based on the kinematical age of the system as derived from its
$(U,V,W)$ space motion and the absence of H$\alpha$ emission from the
Gl~570BC M-type binary.  The former is a statistical measure and
therefore is not a very strong constraint for an individual star.  The
latter is also not a strong constraint, since the age dependence of
H$\alpha$ emission for field M~dwarfs is also largely based on kinematic
analyses \citep{1996AJ....112.2799H}.  The $\approx$1~Gyr lower age
limit discussed here based on stellar activity indicators for the K-type
primary Gl~570A is a more conservative estimate, especially since these
indicators are calibrated with data from open
clusters.\footnote{\citet{2006ApJ...647..552S} suggest an age range of
  3--5 Gyr for the Gl~570ABCD system, based on modeling the optical to
  mid-IR spectrum of the T-dwarf companion \globject.}

Assuming that the brown dwarfs are coeval with their primary stars,
Figure~\ref{fig:physical-parms} illustrates the model-derived masses,
surface gravities, and effective temperatures for \hdobject\ and
\globject, based on the \citet{1997ApJ...491..856B} evolutionary models
and the observational constraints in Table~1.  The \Lbol\ measurements
more strongly constrain \Teff, while the age of the primary star sets
\logg.
For \globject\ and a nominal age of 2~Gyr, the bolometric luminosity
of 10$^{-5.53\pm0.05}$~\Lsun\ \citep{gol04} gives an effective
temperature of 780~K and a mass of 33~\Mjup, with ranges of 750--825~K 
and 24--51~\Mjup\ for ages of 1--5~Gyr.
(Note that the apparent overlap of the uncertainties in the
Figure~\ref{fig:physical-parms} is misleading, since the uncertainties
represent the plausible spread in the absolute ages.  The relative ages
of the primary stars are known to better accuracy, as just discussed.)


\subsection{Surface Gravity and Metallicity Effects in Late-T Dwarf
Spectra} 

Given the age and metallicity determinations for the primary stars,
\hdobject\ and \globject\ provide two benchmarks for examining the
differential effects of surface gravity and metallicity on the spectra
of late-T dwarfs.
Figure~\ref{fig:physical-parms} shows that the older age of
\hdobject\ leads to a $\approx$0.3~dex higher inferred surface gravity
compared to \globject, with nearly the same \Teff\ for the two objects.
For late-T dwarfs, this difference would be most prominently manifested
in the $K$-band emission, which is heavily influenced by opacity from
collision-induced \htwo\ absorption \citep[e.g.][]{1969ApJ...156..989L}.
Higher surface gravity objects will have higher photospheric pressures,
and thus the \htwo\ opacity will be stronger and the $K$-band flux more
heavily depressed.  Indeed, variations in the near-IR properties of
late-T dwarfs are typically attributed to variations in surface gravity
\citep[e.g.][]{2004AJ....127.3553K, 2006ApJ...639.1095B,
  2005astro.ph..9066B}, which translates into variations in mass because
the radii of field brown dwarfs are very similar.

However, the inferred surface gravity difference does not seem to be in
accord with the appearance of the near-IR spectra of these two objects.
Since \hdobject\ and \globject\ have almost identical metallicities
($\Delta$[Fe/H] $\approx0.1$~dex) and effective temperatures
($\Delta{\Teff}\approx30$~K), the higher surface gravity of
\hdobject\ should lead to bluer near-infrared colors compared to
\globject\ due to stronger \htwo\ opacity.  But instead the observed
colors of \hdobject\ are {\em redder} by 0.12$\pm$0.08 and
0.15$\pm$0.08~mag at $H-K$ and $J-K$, respectively.  Thus, there appears
to be a discrepancy between the expected behavior of \htwo\ opacity and
the actual \hdobject/\globject\ comparison.
Similarly, \citet{2006ApJ...639.1095B} use the pressure sensitivity of
\htwo\ opacity to constrain the surface gravity of T~dwarfs through
spectral indices that ratios the $K$ and $H$-band peak fluxes.  Their
Figure~5 implies that increasing gravity by 0.3~dex at $T_{\rm
  eff}\approx800$~K should produce a decrease in the $K/H$ ratios of
$\approx$20\% --- instead we see an {\em increase} of 17$\pm$8\%
(Table~\ref{table:bbk-indices}).
In short, comparing the near-IR colors and spectra of \hdobject\ and
\globject\ suggest that \hdobject\ has a lower surface gravity and thus
a younger age, in contradiction to the relative ages inferred for their
primary stars in \S~4.1.

We suggest that this apparent discrepancy arises from the small
metallicity difference ($\approx$0.1~dex) between the two systems.
Collision-induced $\htwo$ opacity is also expected to be affected by
metallicity variations \citep{1997A&A...324..185B}.  Higher
metallicities lead to more opaque atmospheres; therefore, the
photosphere resides at lower pressures and collision-induced
\htwo\ opacity is reduced.
Models of T~dwarf spectra by \citet{2005astro.ph..9066B} as a function
of metallicity ([Fe/H] = $-$0.5 to +0.5) and surface gravity (\logg =
4.5 to 5.5) show that both variables can have strong effects on the
emergent near-IR SED.  For instance, Figure~20 of Burrows \etal\ shows
that at fixed \Teff, the $J-K$ model colors vary greatly with
metallicity and surface gravity and that the two quantities act in
opposite senses, as expected: the higher surface gravities that produce
bluer near-IR colors can be counteracted by higher metallicities leading
to redder colors.
Likewise, models by M. Marley \etal\ (in prep.) suggest that the
$\approx$0.1~dex metallicity difference between \hdobject\ and
\globject\ could counteract and even outweigh the effect of the 0.3~dex
difference in \logg\ on their relative $K$-band fluxes.


To quantify the sensitivity of near-IR spectra to changes in metallicity
and surface gravity, we examine the condensate-free atmospheric models
from the Tucson group, as described in \citet{2002ApJ...573..394B} and
\citet{2005astro.ph..9066B}.  \citet{2006ApJ...639.1095B} define a set
of 5 spectral indices to characterize the emission peaks and
\htwoo\ absorption in late-T dwarf spectra; our measurements for
\hdobject\ and \globject\ are given in Table~\ref{table:bbk-indices}.
For a particular spectral index, one can envision that its
model-predicted values constitute a 3-dimensional surface in the
parameter space of \{\Teff, \logg, $Z$\}.  We compute the local slope of
the surface to quantify how the index varies with effective temperature,
surface gravity, and metallicity about nominal reference values of
$\Teff = 800$~K, $\log(g)=5.0$ and $Z/\Zsun=1.0$.  In practice, the
indices do not vary linearly, and thus we compute the changes both for
decreasing and increasing values of \Teff, \logg, and $Z$ about the
reference values.  For example, the \htwoo-$J$ and \htwoo-$H$ indices
from the models change strongly from $Z/\Zsun=0.3$ to 1.0 but not very
much for $Z/\Zsun=1.0$ to 3.0.

Figure~\ref{fig:compute-ratios-gradient} provides a pictoral
representation of the metallicity and surface gravity sensitivities for
\Teff\ = 700 to 900~K.  The results for the 800~K models are also listed
in Table~\ref{table:sensitivities}.
The models predict that indices involving the $K$-band flux, namely
$K/H$ and $K/J$, are strongly sensitive to both metallicity and surface
gravity, and less sensitive to \Teff.
Table~\ref{table:sensitivities} indicates that a 0.1~dex greater
metallicity for \hdobject\ compared to \globject\ will produce a $2.41
\times 0.1 = 24\%$ increase in $K/H$, compared to the $-0.58 \times 0.3
= 17\%$ decrease due to a 0.3~dex higher surface gravity, consistent
with our hypothesis that that the enhanced $K$-band flux of
\hdobject\ is due to its higher metallicity, despite its higher surface
gravity.\footnote{The difficulties of disentangling surface gravity and
  metallicity effects are also demonstrated by
  \citet{2006astro.ph..9793L}.  Their analysis of the unusual strong
  H$\alpha$-emitting T6.5~dwarf 2MASS~1237+6526 highlights the
  degeneracy betwen inferring subsolar metallicity and reduced surface
  gravity from low-resolution near-IR spectra.}  For comparison,
  Table~\ref{table:sensitivities} also indicates that a 30~K difference
  in \Teff\ leads to only a $0.21 \times (30/100) = 6\%$ change
  in $K/H$.

But are the model predictions correct?  A comparison of \hdobject\ and
\globject\ allows, for the first time, an empirical test of the {\em
  differential} effects of \logg\ and metallicity on late-T dwarf
spectra.  Based on the properties compiled in Table~1, we adopt a
difference of 0.25$\pm$0.05~dex in surface gravity, 0.10$\pm$0.05~dex in
metallicity, and 20$\pm$10~K in temperature between these two objects,
with all the quantities being larger for \hdobject.
Table~\ref{table:bbk-indices} summarizes the results.  The model
predictions for $K/H$ agree best with the observations given the
uncertainties, showing the index increases for \hdobject\ compared to
\globject.  The other model results do not agree so well with the
observations, with those involving the $J$-band flux being particularly
poor ($Y/J$ and $K/J$), and is likely affected by missing \meth\ and
NH$_3$ opacities in the $YJH$-bands (\eg, Figure~15 of \citealp{bur01}).
The $Y$-band fluxes are also subject to uncertainties in the far-red
opacity wings of the \ion{K}{1}~0.77~\micron\ resonance doublet; the
Tucson models employed here are based on the modified Lorentzian
profiles from \citet{1999ApJ...512..843B}, as opposed to more recent
calculations by \citet{2003ApJ...583..985B}.

\citet{2006ApJ...639.1095B} have used the same Burrows \etal\ models,
largely restricted to solar metallicity, to determine \Teff\ and
\logg\ for late-T dwarfs from near-IR spectra and, combined with
evolutionary models, to estimate ages and masses.  In principle, such a
method could be quite valuable, as the key physical parameters of the
coldest known brown dwarfs could be extracted solely from low-resolution
near-IR spectra.  They calibrate their method by scaling the models to
agree with the \Teff\ and \logg\ determined for \globject\ by
\citet{2001ApJ...556..373G}.  While such a calibration is necessary, it
is not complete, since it only requires the models to agree with a
single object.  \hdobject\ now provides second calibration point, and
our differential comparison indicates that (1) non-solar metallicities
must be considered to determine the correct \logg\ (and thus the mass),
and (2) model predictions for indices involving the $J$-band flux may
not be correct.  Improvements in the models and identification of more
calibration objects will be needed to fully exploit the potential of the
\citet{2006ApJ...639.1095B}
approach.\footnote{\citet{2006astro.ph..9556B} have applied this method
  to near-IR spectra of \hdobject.  Similar to us, they find that
  accounting for metallicity is important in extracting the physical
  parameters.  However, they infer a \logg\ and age similar to
  \globject, even after accounting for the super-solar metallicity of
  \hdobject.  This is in discord with our conclusion that the
  \hdobject\ is $\approx$3$\times$ older than \globject\ and thus has a
  higher surface gravity.}

Our comparison here of the near-IR colors and low-resolution spectra of
\hdobject\ and \globject\ indicates that metallicity effects are
significant in late-T dwarf spectra, perhaps more important than
previously appreciated.
If this indeed is the case, then disentangling the effects of gravity
and metallicity variations among the field population may be quite
challenging.  In a similar vein, analyses that consider only solar
metallicity models \citep[e.g.][]{2004AJ....127.3553K,
  2006ApJ...639.1095B} will naturally infer an inflated spread in
\logg\ for the late-T dwarfs compared to the true spread.  Since field
brown dwarfs have very similar radii, errors in \logg\ determinations
translate almost directly into errors in the masses.

We provide here a simple estimate of the relative importance of
metallicity versus surface gravity variations.  At \Teff~=~700--900~K,
objects with ages of 1--10~Gyr correspond to masses of 20--65~\Mjup,
based on \citet{1997ApJ...491..856B} models. Given the nearly constant
radii of old substellar objects, this age spread amounts to about a
0.5~dex variation in surface gravity.  The metallicity spread of the
field population is about the same; for instance,
\citet{1996MNRAS.279..447R} find that $\approx$90\% of the G~dwarfs in
the solar neighborhood cover a metallicity spread of $\approx$0.6~dex.
For comparison, Figure~\ref{fig:compute-ratios-gradient} and
Table~\ref{table:sensitivities} indicate that the $K/H$ index changes by
100--240\% per dex of $\log(Z)$ change and by 60--100\% per dex of
\logg\ change, \ie, that $K/H$ is about twice as sensitive to
metallicity as compared to surface gravity.  Given the similar spread in
\logg\ and $\log(Z)$ in the field population, the models therefore
predict that metallicity variations will have {\em a larger effect} than
surface gravity variations on the spectral properties of the late-T
dwarfs.\footnote{In fact, the empirical comparison of \hdobject\ to
  \globject\ suggests an even larger effect due to metallicity than the
  models do.  Table~\ref{table:bbk-indices} shows that $K/H$ changes by
  almost twice as much between these two objects as the models predict.}


\section{Conclusions}

We have obtained near-IR photometry and spectroscopy of \hdobject, the
wide-separation, low-luminosity companion to the nearby K0V star
HD~3651A identified by \citet{2006astro.ph..8484M} and
\citet{2006astro.ph..9464L}.  We find a spectral type of T7.5$\pm$0.5
for \hdobject.  This makes it the 6th T~dwarf with a spectral type later
than T7 and only the 4th with a trigonometric parallax.  It is also the
6th T dwarf found as a companion, the other ones being Gl~229B (T7p;
\citealp{1995Natur.378..463N}), Gl~570D (T7.5;
\citealp{2000ApJ...531L..57B}), $\epsilon$~Ind~Bab (T1+T6 binary;
\citealp{2004A&A...413.1029M}), and SCR~1845$-$6357B ($\approx$T5.5;
\citealp{2006ApJ...641L.141B}).  We find that \Teff\ and \Lbol\ for
\hdobject\ are among the lowest determined for any brown dwarf with a
trigonometric parallax, comparable to those of the T7.5~dwarf Gl~570D and
exceeded only by the T8 dwarf 2MASS~0415$-$0935.

HD~3651B appears to be very similar to \globject, the other known late-T
dwarf companion to a K~dwarf.
Given their similar temperatures ($\Delta\Teff\approx30$~K) and the
similar metallicities of their host stars
($\Delta$[Fe/H]$\approx0.1$~dex), a comparison of \hdobject\ to
\globject\ offers a probe of surface gravity effects in ultracool
atmospheres.  While absolute age determinations are difficult for field
dwarfs, several methods indicate that HD~3651A (3--12~Gyr) is notably
($\approx$3$\times$) older than Gl~570A (1--5~Gyr), and consequently
evolutionary models show that \hdobject\ has a $\approx$0.3~dex higher
surface gravity than \globject.
However, the near-IR colors and spectra of \hdobject\ compared to
\globject\ appear to be at variance with the relative ages of the host
stars, with the $K$-band fluxes indicating a lower surface gravity, and
thus younger age, for \hdobject.  Hence, at face value the relative ages
these two brown dwarfs derived from activity-based indicators for
primary stars and from the interpretation of late-T~dwarf spectra appear
to be in discord.

We suggest that this discrepancy arises from the small
($\approx$0.1~dex) metallicity difference between the two objects,
reflecting the metallicity sensitivity of the collision-induced
\htwo\ opacity that shapes the $K$-band flux.  
Given the physical properties established from their host stars,
\hdobject\ and \globject\ offer a differential test of gravity and
metallicity effects in late-T dwarf spectra.  We find that the the
condensate-free models of \citet{2005astro.ph..9066B} are in fair
agreement with the observed change in $K/H$ index but other indices,
especially those involving the $J$-band flux, do not match the
observations very well.
Simple estimates based on the plausible metallicity and gravity spread
among the field population suggest that metallicity may be more
important than surface gravity in producing spectral variations among
late-T dwarfs.
Hence, theoretical atmospheres that include non-solar metallicities will
be valuable for interpreting these ultracool objects.

\begin{figure}
\hskip -0.4in
\centerline{\includegraphics[width=5in,angle=90]{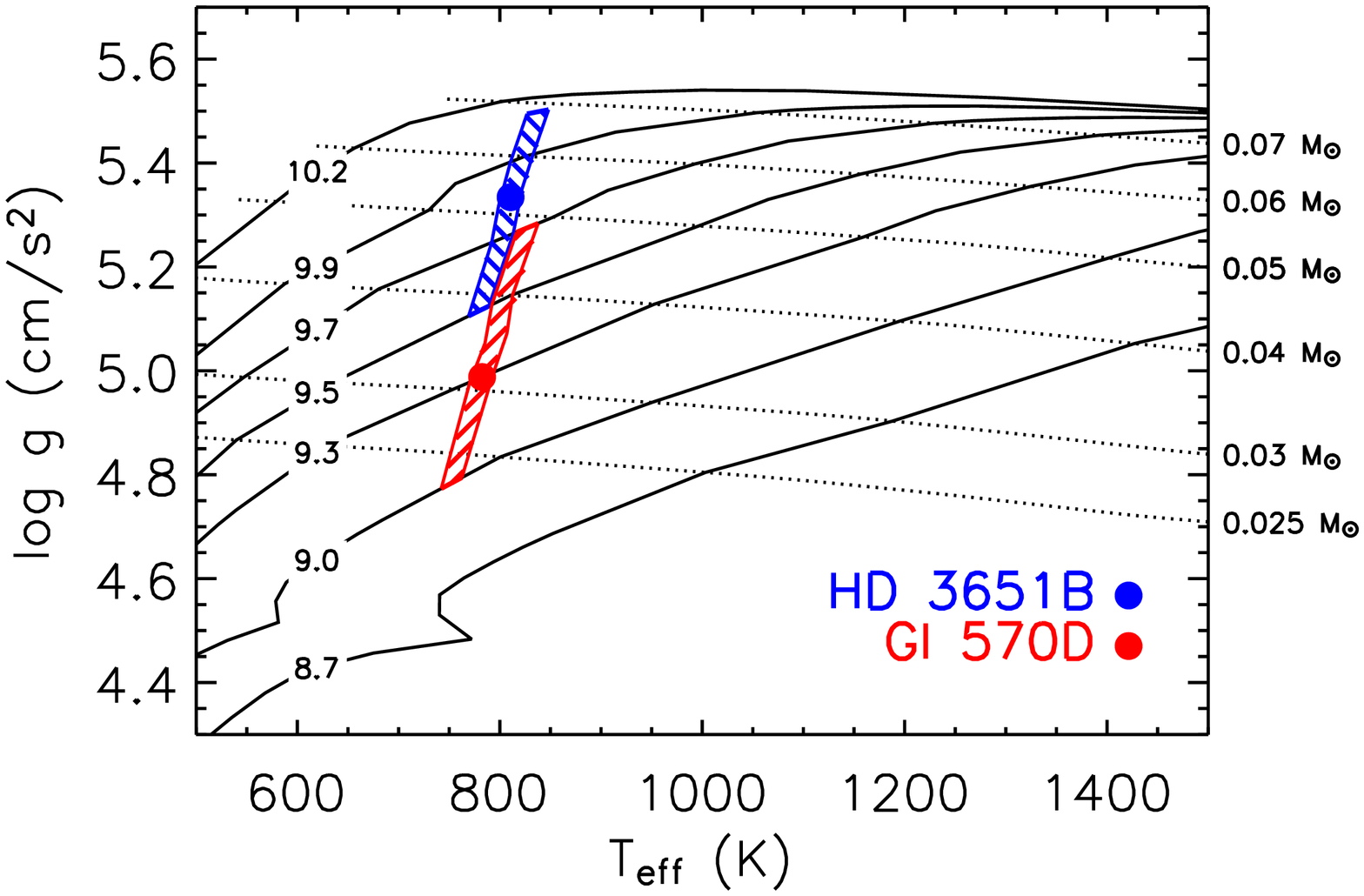}}
\vskip -4ex
\caption{\normalsize Inferred \Teff\ and \logg\ for \hdobject\ and
  \globject\ based on sol,ar-metallicity evolutionary models by
  \citet{1997ApJ...491..856B} and observational constraints summarized
  in Table~1.  Solid lines are isochrones from 0.5 to 16~Gyr, labeled by
  the logarithm of their age.  Dotted lines represent iso-mass models,
  labeled on the right side of the plot in units of \Msun.  The solid
  colored dots indicate the model-derived \Teff\ and \logg\ for
  \hdobject\ and \globject, given the computed \Lbol's and estimated
  ages of the objects.  The \Teff\ of the two objects are very similar,
  with \hdobject\ having a $\approx$0.3~dex higher surface gravity.  The
  colored hatched regions indicate the observational uncertainties.
  Note that the apparent overlap of the uncertainties in determining
  \logg\ is somewhat misleading, since each colored region represents
  the uncertainty in the absolute ages; the relative ages of the primary
  stars is better constrained, with \hdobject\ being $\approx$3$\times$
  older than \globject\ (\S~4.1). \label{fig:physical-parms}}
\end{figure}

\begin{figure}
\hskip 0.15in
\includegraphics[width=4.5in,angle=90]{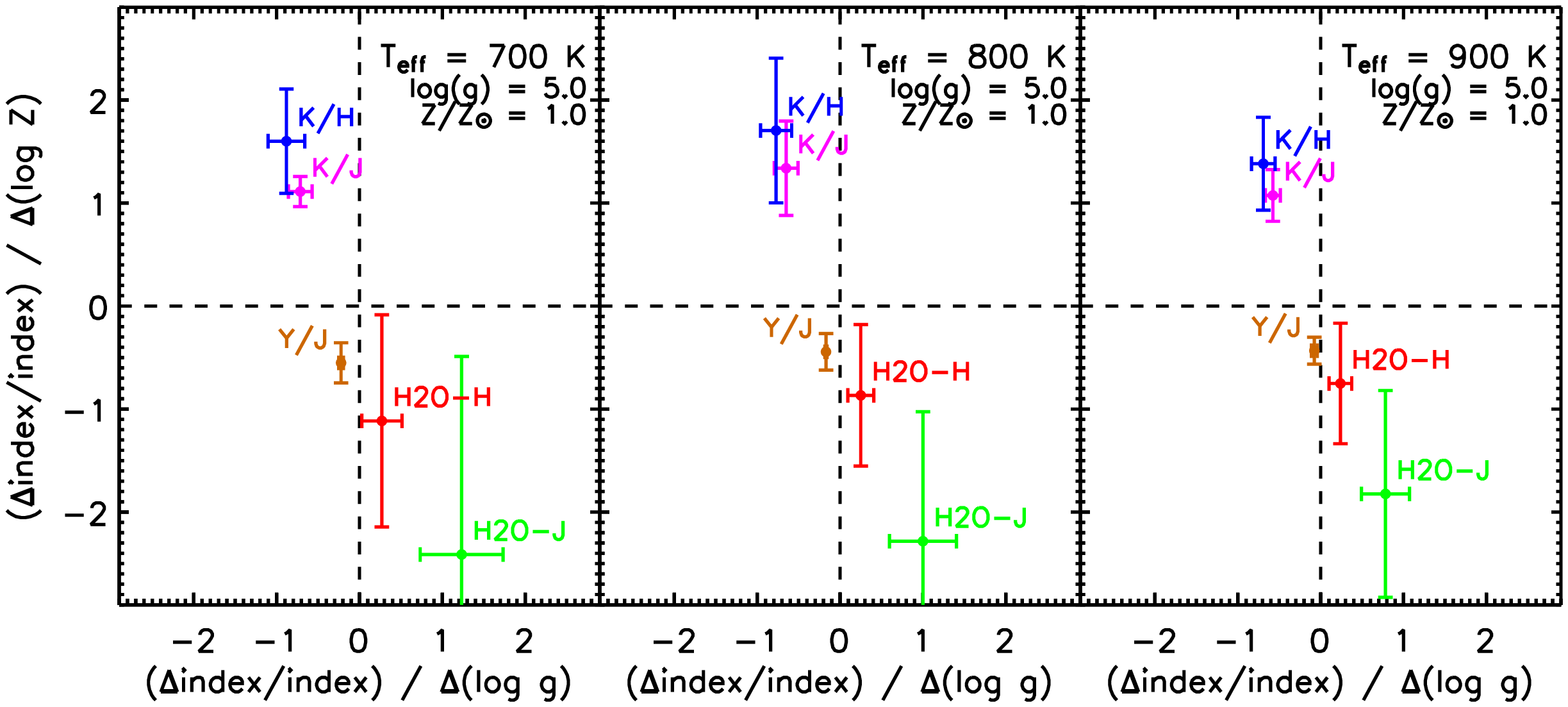}
\vskip -8ex
\caption{\normalsize Quantitative representation of the metallicity and
  surface gravity sensitivity of the \citet{2006ApJ...639.1095B}
  spectral indices, based on condensate-free atmosphere models by
  \citet{2005astro.ph..9066B}.  The models have been smoothed to a
  spectral resolution of 100~\AA, appropriate for IRTF/Spex data (though
  the results are essentially the same for 10$\times$ higher spectral
  resolution).  For each spectral index, the $x$-axis value gives its
  fractional change as \logg\ is changed and the $y$-axis as $Z$ is
  changed, about a nominal reference value of $\logg = 5.0$ and
  $Z/\Zsun=1.0$.  A purely metallicity-sensitive index would have a
  large ordinate value but an abscissa of zero (\ie, it would lie on the
  vertical line), while a purely gravity-sensitive index would lie on
  the horizontal line.  The error bars show the range in the sensitivity
  of the indices as \logg\ changes over the intervals [4.5, 5.0] and
  [5.0, 5.5] and as $Z/\Zsun$ changes over the intervals [0.3, 1.0] and
  [1.0, 3.0].  This plot indicates that the $K/H$ and $K/J$ indices are
  very sensitive to both changes in metallicity and surface gravity at
  fixed \Teff. \label{fig:compute-ratios-gradient}}
\end{figure}

Further observations of the HD~3651AB and Gl~570ABCD system 
will clarify the relative roles of metallicity and surface gravity in
late-T~dwarf spectra.  More accurate constraints on the ages of the
primary stars would better constrain the surface gravities (\eg,
Figure~\ref{fig:physical-parms}).  Higher resolution near-IR spectra,
\eg, to resolve the $J$-band K~I 1.25~\micron\ doublet, might probe
gravity and metallicity diagnostics, as could spectra at far-red optical
wavelengths \citep[e.g.][]{2002ApJ...573..394B, 2003ApJ...594..510B}.
Discovery of more late-T dwarfs as stellar companions and/or members of
nearby star clusters will provide more benchmarks for comparative study
over a range of effective temperatures, surface gravities, and
metallicities.

Finally, confirmation of the substellar status of
\hdobject\ demonstrates that wide-separation brown dwarf companions and
giant planets can co-exist around the same star.
While brown dwarf companions have been resolved around other stars
\citep[e.g.][]{1988Natur.336..656B, 1995Natur.378..463N,
  1998Sci...282.1309R, 2001AJ....121.3235K, 2001astro.ph.12407L,
  2004ApJ...617.1330M},
and there are also several systems with both brown dwarf and planetary
companions found by radial velocity searches
\citep[e.g.][]{2001ApJ...555..418M, 2002astro.ph..2458U}, the HD~3651ABb
system is the first with a radial velocity planet and a wide-separation
(resolved) brown dwarf.
The upcoming Pan-STARRS project \citep{2002SPIE.4836..154K} will monitor
the entire sky observable from Hawai`i and produce sensitive multi-band
optical photometry with high astrometric precision.  This should be
promising means to identify more wide-separation substellar companions
to stars by their common parallax and proper motion, including stars
being monitored by high precision radial velocity surveys.  Therefore,
as our census of the solar neighborhood continues to become more
complete, so will our appreciation for the diversity of planetary
systems.


\acknowledgments

It is a pleasure to thank Paul Hirst, Tim Carroll, John Rayner, Bill
Golisch, Alan Tokunaga, and the staff of UKIRT and IRTF for their
assistance with the observations.  We thank Adam Burrows, Kevin Luhman,
and Adam Burgasser for providing digital versions of published results,
and thank John Rayner and Michael Cushing for assistance with SpeXtool.
We are grateful to our anonymous referee, Adam Burrows, and Katelyn
Allers for useful comments.  Our research has benefitted from the 2MASS
data products; NASA's Astrophysical Data System; and the SIMBAD database
operated at CDS, Strasbourg, France.
MCL acknowledges support for this work from NSF grants AST-0407441 and
AST-0507833 and an Alfred P. Sloan Research Fellowship.  SKL is
supported by the Gemini Observatory, which is operated by the
Association of Universities for Research in Astronomy, Inc., on behalf
of the international Gemini partnership of Argentina, Australia, Brazil,
Canada, Chile, the United Kingdom, and the United States of America.
The United Kingdom Infrared Telescope is operated by the Joint Astronomy
Centre on behalf of the U.K. Particle Physics and Astronomy Research
Council.
Finally, the authors wish to recognize and acknowledge the very
significant cultural role and reverence that the summit of Mauna Kea has
always had within the indigenous Hawaiian community.  We are most
fortunate to have the opportunity to conduct observations from this
mountain.

\clearpage



\clearpage

\begin{deluxetable}{lcc}
\tablecaption{Properties of the Late-T Dwarf Companions \hdobject\ and \globject} 
\tablewidth{0pt}
\tablehead{
  \colhead{Property} &
  \colhead{\hdobject} &
  \colhead{\globject}
}

\startdata
Spectral type               &  T7.5 $\pm$ 0.5                     & T7.5 $\pm$ 0.5\tablenotemark{a} \\
Primary star spectral type  & K0V\tablenotemark{b}                & K4V\tablenotemark{b} \\
Distance (pc)               &  11.11 $\pm$ 0.09\tablenotemark{c}  & 5.90 $\pm$ 0.06\tablenotemark{c} \\
Estimated age (Gyr)         &  6$_{-3}^{+6}$\tablenotemark{d}      & 2$_{-1}^{+3}$\tablenotemark{d} \\
$[$Fe/H$]$                  &  0.09 -- 0.16\tablenotemark{e}      & 0.01 -- 0.10\tablenotemark{e} \\
$J$ (mags)                  &  16.31 $\pm$ 0.03                   & 14.82 $\pm$ 0.05\tablenotemark{f} \\
$H$ (mags)                  &  16.72 $\pm$ 0.03                   & 15.28 $\pm$ 0.05\tablenotemark{f} \\
$K$ (mags)                  &  16.86 $\pm$ 0.03                   & 15.52 $\pm$ 0.05\tablenotemark{f}\\
$J-H$                       &  $-$0.41 $\pm$ 0.04                 & $-$0.46 $\pm$ 0.07\tablenotemark{f} \\
$H-K$                       &  $-$0.14 $\pm$ 0.04                 & $-$0.24 $\pm$ 0.07\tablenotemark{f} \\
$J-K$                       &  $-$0.55 $\pm$ 0.04                 & $-$0.70 $\pm$ 0.07\tablenotemark{f} \\
$M_{J}$ (mags)              &  16.08 $\pm$ 0.03                   & 15.97 $\pm$ 0.05\tablenotemark{f} \\
$M_{H}$ (mags)              &  16.49 $\pm$ 0.03                   & 16.43 $\pm$ 0.05\tablenotemark{f} \\
$M_{K}$ (mags)              &  16.63 $\pm$ 0.03                   & 16.67 $\pm$ 0.05\tablenotemark{f} \\
$\log(\Lbol/\Lsun)$         & $-$5.58 $\pm$ 0.05                  & $-$5.53 $\pm$ 0.05\tablenotemark{g} \\
Mass (\Mjup)                & 56$_{-16}^{+16}$                     & 33$_{-9}^{+18}$ \\
\Teff\ (K)                  & 810$_{-30}^{+30}$                   & 780$_{-30}^{+45}$  \\
$\log(g)$ (cgs)             & 5.3$_{-0.2}^{+0.2}$                 & 5.0$_{-0.2}^{+0.3}$ \\
\enddata

\tablecomments{The tabulated results are from this paper, unless
  otherwise noted.  All photometry is on the MKO system.}

\tablenotetext{a}{\citet{2005astro.ph.10090B}}
\tablenotetext{b}{From SIMBAD.}
\tablenotetext{c}{Parallax for the primary star from \citet{1997A&A...323L..49P}.} 
\tablenotetext{d}{See \S~3.2, \S~4.1, and Table~3.}
\tablenotetext{e}{\citet{1997A&AS..124..299C, 1998A&AS..129..237F,
  2003AJ....126.2048G, 2004A&A...415.1153S, 2005ApJS..159..141V,
  2005A&A...437.1127S}}
\tablenotetext{f}{\citet{2001ApJ...556..373G}}
\tablenotetext{g}{\citet{gol04}}

\end{deluxetable}


\begin{deluxetable}{lccccc}
\tablecaption{Spectral Typing of \hdobject}
\tablewidth{0pt}
\tablehead{
  \colhead{Dataset} &
  \colhead{\htwoo-$J$} &
  \colhead{\meth-$J$}  &
  \colhead{\htwoo-$H$} &
  \colhead{\meth-$H$}  &
  \colhead{\meth-$K$}
}

\startdata
UKIRT/CGS4 &  \ldots    & 0.214 (T7.5) & 0.212 (T7.5) & 0.128 (T7.5) & 0.082 ($\ge$T7) \\
IRTF/Spex  & 0.058 (T8) & 0.192 (T8)   & 0.197 (T7.5) & 0.121 (T8)   & 0.059 ($\ge$T7) \\
\enddata

\tablecomments{Measurements of spectral indices along with the
  corresponding spectral type in parentheses based on the
  \citet{2005astro.ph.10090B} classification scheme.}

\end{deluxetable}

\begin{deluxetable}{lcccc}
\tablecaption{Age Estimates for HD~3651A and Gl~570A}
\tablewidth{0pt}
\tablehead{
  \colhead{Method} &
  \colhead{Ref}    &
  \colhead{HD 3651A} & 
  \colhead{Gl 570A}  &
  \colhead{Relative age} \\
  \colhead{} &
  \colhead{}    &
  \colhead{(Gyr)} & 
  \colhead{(Gyr)}  &
  \colhead{HD 3651A / Gl 570A} 
}

\startdata
CaII HK emission &  1,2  &   5--9       &  0.8--2.2          & $\approx$ 5   \\
X-ray emission   &  3    & 0.9--2.2     &  0.4--0.8          & $\approx$ 2.5 \\
Rotation period  &  4,5  &    15        &      6             & 2.5 \\    
Isochrones       &  6    & 8$_{-5}^{+4}$ &  3.3$_{-3.1}^{+8.3}$ & $\approx$ 2.5 \\
Isochrones       &  7    & $>$11.8      &  $<$0.6            & $>$20        \\
\enddata

\tablecomments{Age estimates for each star are discussed in \S~3.2 and
  \S~4.1.  The relationship used to convert the observed quantity (\eg,
  X-ray emission) to the stellar age is given in the cited reference.
  In computing the relative ages in the last column, the geometric mean
  is used for methods that produce an age range.}

\tablerefs{
(1) \citet{don93};
(2) \citet{1998csss...10.1235D};
(3) \citet{1998PASP..110.1259G};
(4) \citet{1972ApJ...171..565S};
(5) \citet{1999A&A...348..897L}; 
(6) \citet{2005ApJS..159..141V};
(7) \citet{2006astro.ph..7235T}
}

\end{deluxetable}


\begin{deluxetable}{lcccc}
\tablecaption{Spectral Index Measurements for \globject\ and
  \hdobject \label{table:bbk-indices} }
\tablewidth{0pt}
\tablehead{
  \colhead{Index} &
  \colhead{\globject} &
  \multicolumn{2}{c}{\hdobject} \\
  \colhead{}  &
  \colhead{Observed}  &
  \colhead{Observed} &
  \colhead{Predicted} 
}

\startdata
\htwoo-$J$ &  0.066 $\pm$ 0.005  &  0.058 ($\pm$ 0.003) &  0.087$_{-0.006}^{+0.006}$ \\
\htwoo-$H$ &  0.204 $\pm$ 0.005  &  0.205 $\pm $ 0.011  &  0.237$_{-0.010}^{+0.010}$ \\
$Y/J$      &  0.412 $\pm$ 0.025  &  0.449 ($\pm$ 0.027) &  0.387$_{-0.007}^{+0.007}$ \\
$K/J$      &  0.110 $\pm$ 0.008  &  0.142 $\pm $ 0.008  &  0.116$_{-0.013}^{+0.014}$ \\
$K/H$      &  0.250 $\pm$ 0.005  &  0.292 $\pm $ 0.022  &  0.276$_{-0.040}^{+0.043}$ \\
\enddata

\tablecomments{Measurements of \citet{2006ApJ...639.1095B} spectral
  indices using our spectra for \hdobject\ and spectra from
  \citet{2001ApJ...556..373G} and \citet{2006ApJ...639.1095B} for
  \globject.  For the second and third columns, the error bars are
  computed from the standard deviation of the UKIRT/CGS4 and IRTF/Spex
  measurements.  Where only Spex data are available, we assume an error
  of 6\%, based on the median errors of the other indices, and place it
  in parentheses. The last column gives the model-predicted values for
  \hdobject, assuming a 0.10$\pm$0.05~dex higher metallicity, a
  0.25$\pm$0.05~dex higher surface gravity, and 20$\pm$10~K hotter
  temperature than \globject.  See \S~4.2 for a full explanation.}

\end{deluxetable}


\begin{deluxetable}{lcccccc}
\tablecaption{Sensitivity of Spectral Indices to Surface Gravity,
  Metallicity, and Effective Temperature for \Teff\ = 800 K, \logg = 5.0
  (cgs), and $Z/\Zsun=1.0$ \label{table:sensitivities}}
\tabletypesize{\small} \tablewidth{0pt}
\tablehead{
  \colhead{Spectral Index} &
  \multicolumn{2}{c}{($\Delta$(index)/index)/$\Delta$(log $g$)} &
  \multicolumn{2}{c}{($\Delta$(index)/index)/$\Delta$(log $Z$)} &
  \multicolumn{2}{c}{($\Delta$(index)/index)/$\Delta$(\Teff/100~K)} \\
  \colhead{}   & 
  \colhead{[4.5, 5.0]}   & 
  \colhead{[5.0, 5.5]}   & 
  \colhead{[$-$0.5, 0.0]}   & 
  \colhead{[0.0, +0.5]}    &
  \colhead{[700~K, 800~K]} &
  \colhead{[800~K, 900~K]}
}

\startdata
\htwoo-$J$  & \phs0.60 & \phs1.40 & $-$3.54  & $-$1.03  &  0.36  &  0.47  \\
\htwoo-$H$  & \phs0.09 & \phs0.41 & $-$1.55  & $-$0.18  &  0.42  &  0.37  \\
$Y/J$       & $-$0.16  & $-$0.18  & $-$0.62  & $-$0.27  &  0.10  &  0.06  \\
$K/J$       & $-$0.79  & $-$0.51  & \phs0.88 & \phs1.80 &  0.17  &  0.14  \\
$K/H$       & $-$0.96  & $-$0.58  & \phs1.00 & \phs2.41 &  0.22  &  0.21  \\
\enddata

\tablecomments{Dependence of \citet{2005astro.ph.10090B} spectral
  indices as a function of surface gravity, metallicity, and effective
  temperature, computed from condensate-free models by
  \citet{2005astro.ph..9066B} for objects around the reference values of
  \Teff = 800~K and $\log(g)=5.0$ and $Z/\Zsun=1.0$.  The models have
  been smoothed to a spectral resolution of 100~\AA, appropriate for
  IRTF/Spex data, though the results are essentially the same for
  10$\times$ higher spectral resolution ($\lesssim$0.01 difference in
  values).  Each column gives the fractional change in the spectral
  index as the parameters \logg, $Z$, and \Teff\ are changed about the
  reference values, with the change in the parameter listed in the
  second row of headings.  For example, the first row of data shows that
  in the models the \htwoo-$J$ index increases by 0.6 $\times$ 0.5 =
  30\% as \logg\ changes from 4.5 to 5.0, decreases by 3.54 $\times$ 0.5
  = 180\% as $\log(Z)$ changes from $-$0.5 to 0.0, and increases by 0.36
  $\times$ 100/100 = 36\% as \Teff\ changes from 700~K to 800~K.
  Spectral indices with larger positive (negative) values have stronger
  (anti-)correlations with \logg, $Z$, or \Teff.}

\end{deluxetable}

\end{document}